Doping and Defect-Induced Germanene: A Superior Media for Sensing $H_2S$, $SO_2$, and $CO_2$ gas molecules

M. M. Monshi[*], S. M. Aghaei, I. Calizo

Department of Electrical and Computer Engineering, Florida International University, Miami, Florida, USA
*email: mmons021@fiu.edu

**Abstract**

First-principles calculations based on density functional theory (DFT) have been employed to investigate the structural, electronic, and gas-sensing properties of pure, defected, and doped germanene nanosheets. Our calculations have revealed that while a pristine germanene nanosheet adsorbs $CO_2$ weakly, $H_2S$ moderately, and $SO_2$ strongly, the introduction of vacancy defects increases the sensitivity significantly which is promising for future gas-sensing applications. Mulliken population analysis imparts that an appreciable amount of charge transfer occurs between gas molecules and a germanene nanosheet which supports our results for adsorption energies of the systems. The enhancement of the interactions between gas molecules and the germanene nanosheet has been further investigated by density of states. Projected density of states provides detailed insight of the gas molecule's contribution in the gas-sensing system. Additionally, the influences of substituted dopant atoms such as B, N, and Al in the germanene nanosheet have also been considered to study the impact on its gas sensing ability. There was no significant improvement found in doped gas sensing capability of germanene over the vacancy defects, except for $CO_2$ upon adsorption on N-doped germanene.

*Index Terms*- Toxic gas, sensor, vacancy defects, doping, germanene nanosheet

1. **Introduction**

Graphene synthesis spurred the quest to obtain two-dimensional (2D) forms of other materials in group IV of the periodic table as predicted by Takeda and Shiraishi in 1994 [1]. The successful synthesis of silicene in 2012 [2] boosted the surge to obtain germanene, a 2D form of germanium analogous to graphene, bolstered by another prediction in which Cahangirov *et al.* found that germanene appears to be in low-buckled form [3]. Inspired by the theoretical predictions, the fully hydrogenated form of germanene, also known as germanane, was synthesized in 2013 using topochemical deintercalation of the layered calcium digermanide ($CaGe_2$). Dávila *et al.* reported experimental evidence of its synthesis by the dry epitaxial growth of germanene on a Au (111) surface [4]. Graphene-like germanene sheets have also been produced on Pt (111), Ge/Ag (111), and Al (111) substrates and were also epitaxially and mechanically exfoliated using GeH synthesis on $SiO_2$ [5-8].

The honeycomb structure of germanene displays ambipolar character and its charge carriers behave like massless Dirac fermions [3, 9]. Interestingly, the Fermi velocity of germanene is found to be ~$5.6\times10^5$ ms$^{-1}$ and carrier interactions with phonons are 25 times less than those of graphene which may explain its high carrier mobility [10]. Once germanene doped or surface passivated, it offers large magnetic moment and transforms from an AFM to FM state [11, 12]. Even with excellent electronic and magnetic properties, germanene suffers from zero energy band gap. Applying electric field [13], chemisorption of adatom species [14, 15], introducing periodic nanoholes [16], defects [17], doping [11], and edge functionalization are possible means to achieve a band gap in silicene and germanene. Previously, edge functionalization has also been used as a tool for tuning the band gap in nanoribbons [13, 14, 18-23].

Several experimental and theoretical studies have studied graphene related materials for gas sensing [24-29]. Similar to silicene, atoms in germanene are bonded in low buckled form which is mostly instigated by the mixing of sp$^2$ and sp$^3$ hybridization of Ge atoms. It is predicted that germanene may have stronger adsorption of atoms and molecules than graphene due to its buckled structure [14, 30]. Xia *et al.* theoretically investigated common gas molecules adsorption behavior on germanene. They have found that $N_2$, CO, $CO_2$, and $H_2O$ are physisorbed on germanene through van der Waals interactions whereas $NH_3$, NO, $NO_2$ and $O_2$ are chemisorbed through strong covalent bonds [20, 31]. Molecular adsorption on germanene could be used either for gas sensing or band gap engineering of germanene depending on adsorption energy [20]. In our previous study, a significant improvement was found by utilizing Li-functionalized germanene as the adsorbent. Our findings suggest that Li-functionalized germanene shows potential for $CO_2$ capture with a storage capacity of 12.57 mol/kg [32]. Recently, Gupta *et al.* forecasted the possibility of germanene in building new gas sensing material with great stability and sensitivity [33]. Padiha *et al.* investigated the impact of Stone-Wales and vacancies induced electronic and transport properties. They discovered that Stone-Wales and divacancies destroy the dispersion relation near the Fermi level to create scattering centers to reduce the current in germanene [34]. Still, germanene gas sensing capability has not been studied much compared to graphene and silicene. It is proven that defects, doping, and functionalization can certainly increase the surface–adsorbate interaction to improve the sensing activity of the materials [17]. In this study, we are going to investigate doping and defects induced gas sensing capability of germanene using density functional theory (DFT) calculations. The sensing properties of germanene nanosheet would be evaluated by considering adsorption energy, charge transfer, and density of states (DOS).

## 2. Methodology

First-principles DFT calculations combined with nonequilibrium Green's function (NEGF) implemented in Atomistix ToolKit (ATK) package [35-37] are used to analyze band gap, DOS, and adsorption energy of the structures. The exchange-correlation functional uses Generalized Gradient Approximation of Perdew-Burke-Ernzerhof (PBE) to solve Kohn-Sham equations and to expand electronic density. A double-$\zeta$ polarized basis set is adopted. The Grimme vdW correction (DFT-D2) is also undertaken to describe long-range vdW interactions between germanene nanosheet and gas molecules [38, 39]. For this reason, a supplementary term ($E_{vdW}$) is added to the DFT total energy ($E_{DFT}$) to take into considerations of the vdW interactions,

$$E_{DFT-D2} = E_{DFT} + E_{vdW} \tag{1}$$

The freestanding germanene nanosheet is modeled by $4 \times 4 \times 1$ supercell containing 32 Ge atoms. A vacuum space of 40 Å along the z direction in which the structures are not periodic to avoid the possible interactions between the periodically repeated unit cells. The Brillouin zone integration is sampled using Monkhorst–Pack grid of $5 \times 5 \times 1$ $k$-points for structural relaxations and $15 \times 15 \times 1$ $k$-points for total energy, electronic properties, and charge transfer calculations [40]. The density mesh cut-off is set to be 75 Hartree, and the structures are allowed to fully relax until the force on each atom becomes less than 0.01 eV/Å.

First, the adsorption energies of the incident gases on the germanene nanosheet have been evaluated using the following formula

$$E_a = E_{Ge-X} - E_{Ge} - E_X \tag{2}$$

where $X=$ $H_2S$, $SO_2$, and $CO_2$. The first, second, and third terms represent the total energies of the germanene nanosheet with adsorbed gas molecules, pristine germanene nanosheet, and the individual gas molecules, respectively. Secondly, the adsorption energies when gases adsorbed on germanene with vacancy defects and dopants have been determined by the following equation [41]

$$E_f = E_{defected-X/doped-X} - E_{defected/doped-Ge} - E_X \tag{3}$$

where $E_{defected-X/doped-X}$ and $E_{defected-X/doped-X}$ are the total energies for defected/doped germanene with gas adsorbed and without gas adsorbed in the system. Additionally, the formation energies of the vacancy defects have been calculated using the following formula

$$E_f = E_{defected} - \{(x-y)/x\} E_{pristine} \tag{4}$$

where $E_f$ denotes the formation energy, $E_{defected}$ and $E_{pristine}$ are the total energies of the defected and pristine germanene, correspondingly, and $x$ and $y$ are the number of atoms in the germanene nanosheet and the number of atoms eliminated from the nanosheet, respectively.

## 3. Results and Discussion

The optimized structure of a pristine germanene sheet is shown in Figure 1(a). The calculated lattice constant (4.20 Å), Ge-Ge bond length (2.46 Å), and buckling distance (0.727 Å) found in optimized geometry are in good agreement with the previous findings [13, 42, 43]. We have investigated three toxic gases ($H_2S$, $SO_2$, and $CO_2$) to bind with monolayer germanene once exposed. To start the relaxation, the gas molecules can be placed at valley, hill, bridge and hollow sites as shown in Figure 1. Gas molecules can be adsorbed at these specific sites horizontally, vertically or in a slanted form. We started with $H_2S$ adsorption on germanene monolayer, which is one of the toxic gases produced from industrial waste and extremely lethal at concentrations >250 ppm [44]. The most stable configuration of $H_2S$ molecule adsorbed on germanene is shown in Figure 2(a), where $H_2S$ is aligned parallel to the surface of germanene, and the S atom is pointing towards the germanene monolayer at a distance of 1.45 Å. An adsorption energy of -0.57 eV is found for $H_2S$ adsorption on pristine germanene which is better than silicene nanosheet [17] and comparable to silicene nanoribbon [45]. Full structural relaxation shows a slight elongation in the H–S bond from 1.35 to 1.36 Å and reduction in the H–S–H angle from 91.29 to 91.12°. The Ge-Ge bond length varies from 2.45 to 2.49 Å around the $H_2S$ molecule while bond length remains the same (2.46 Å) as pristine (Figure 1(a)) in other locations.

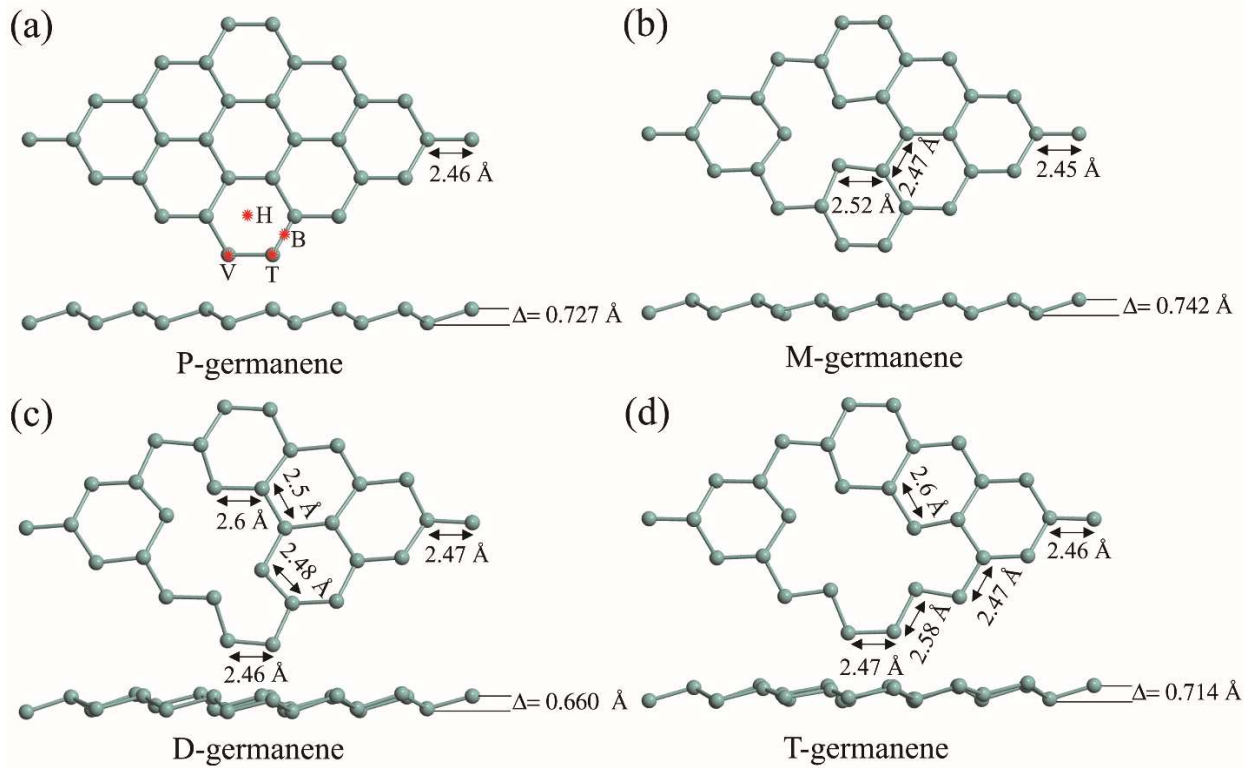

Figure 1. Top and side view of optimized structures of (a) pristine, (b) mono vacancy, (c) di vacancy, and (d) tri vacancy of germanene sheet. Possible sites for gas molecules adsorption on germanene are: T (top), V (Valley), B (bridge), and H (hollow).

Another hazardous sulfur-containing gas, $SO_2$, the prime component of industrial waste materials, has a poisonous effect that is particularly precarious for those with respiratory as well as heart diseases [46]. The minimum energy configuration of $SO_2$ adsorbed germanene has an adsorption energy of −1.55 eV, and the molecule is positioned horizontally at a distance of 2.08 Å from the germanene monolayer, as shown in Figure 2(b). This adsorption energy is sufficiently larger than those obtained for a germanane [33], silicene gas sensor [17] and lower than silicene nanoribbon [45]. S-O bond length stretched to 1.58 from 1.47 Å and the angle reduced to 110.2° from 119.91°. Finally, $CO_2$ adsorption on a germanene nanosheet was also studied. It is the most abundantly available greenhouse gas in the environment. By investigating all of the configurations, it has been found that horizontally adsorbed $CO_2$ is the most stable configuration. This structure provides a tiny adsorption energy of -0.05 eV which is smaller than that of graphene [47], silicene [17] and germanene nanoribbon [20], with a relatively large germanene-$CO_2$ distance of 3.79 Å as can be seen from Figure 2(c). Based these findings, while $SO_2$ is chemisorbed on pristine germanene, H$_2$S is physically adsorbed and CO$_2$ offers fairly weak physisorption that limits the use of pristine monolayer germanene as a gas sensor for CO$_2$ and H$_2$S detection. A possible remedy to this obstacle is to improve the adsorption by the creation of defects and introducing foreign atoms.

In this research, we focused on vacancy defects in order to enhance the sensitivity of germanene nanosheet to the toxic gases. The vacancy defects which are unavoidable in the synthesis of nanomaterials such as exfoliation of monolayers could be created by the exposure from a laser or electron beam thereby enhancing the adsorption mechanism of the adsorbate gas molecules with the host monolayers. This indispensable circumstance inspires us to form mono-, di-, and tri-vacancies in the germanene nanosheet by eliminating one, two and three germanium atoms, correspondingly. For the sake of convenience, we refer mono, di, and trivacancy germanene by M-germanene, D-germanene, and T-germanene, respectively and the pristine one as P-germanene. The optimized atomic structures of M-germanene, D-germanene, and T-germanene are shown in Figure 1(b)-1(d). Stabilities of defects have been determined using equation 4 that are -2.77, -3.70 and -5.03 eV for mono-, di-, and tri vacancies, correspondingly which are very similar for silicene [41]. The calculated energies show that the defected structures are stable because the formation process is exothermic. It is found that the Ge-Ge bond length is elongated due to the introduction of vacancies and buckling for nanosheet changes from 0.727 to 0.742 Å, 0.660 and 0.714 Å for mono, di, and trivacancies, respectively as seen in Figure 1.

The impact of vacancies on gas sensing mechanism has been studied by starting with M-germanene sheet. We have considered various sites to find the most stable configuration for gas molecules to be adsorbed on M-germanene. H$_2$S gas molecules are strongly chemisorbed with an adsorption energy of -2.41 eV which is 5 times higher than its value on the pristine nanosheet. The nanosheet surface-H$_2$S distance considerably reduced to 1.61 Å from 2.95 Å which indicates strong chemisorption and ended up in dissociation of H$_2$S with one of the H atoms dislodges itself. The other H-S bond remains in contact with a distance of 1.4 Å which is larger than its previous value (1.35 Å) as an isolated molecule. The most stable configuration is presented in Figure 2(d). The minimum energy structure for the case of SO$_2$ gas adsorbed on M-germanene (in Figure 2(e)) demonstrates that all of the atoms are attached to Ge atoms in the nanosheet.

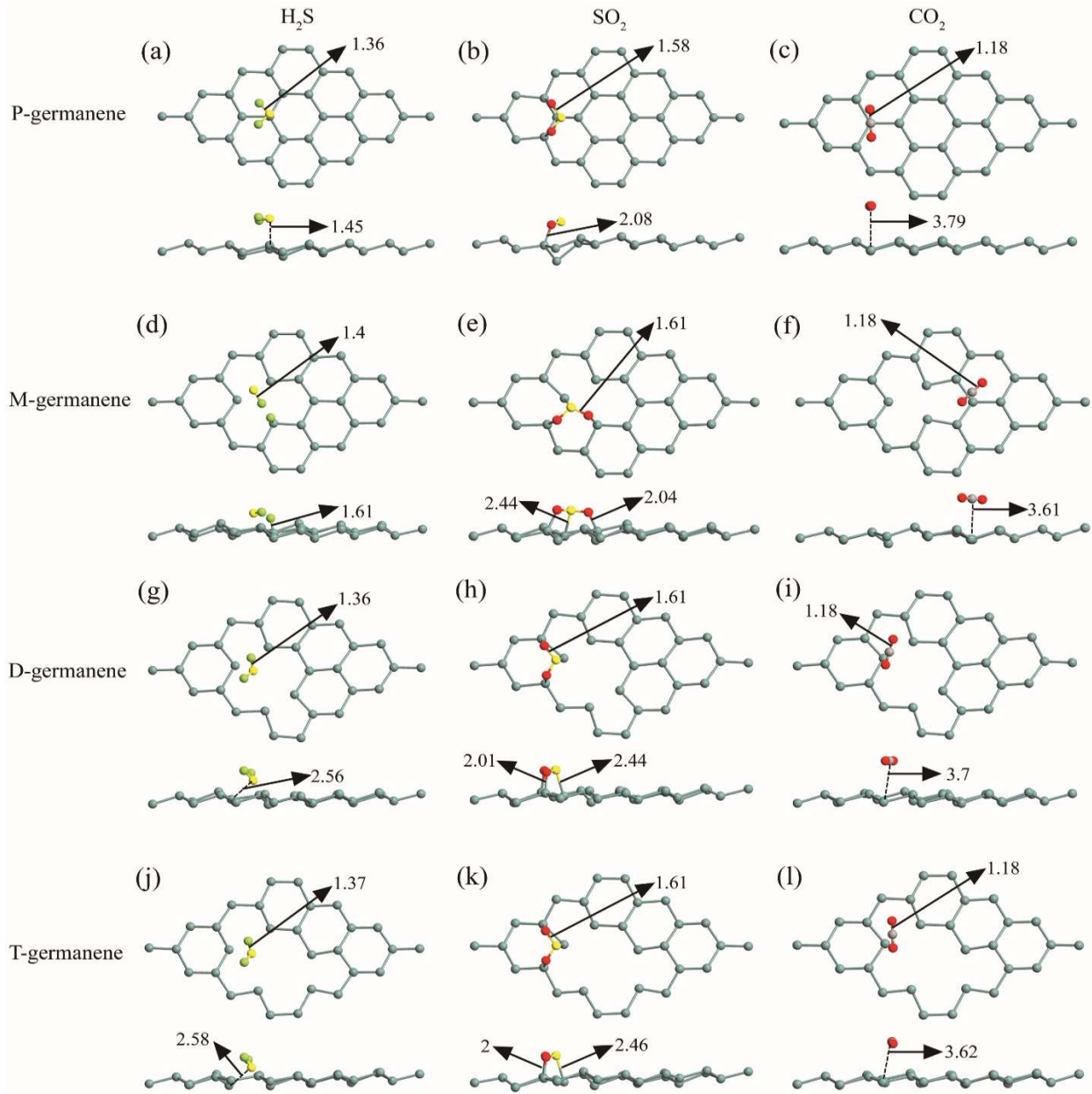

Figure 2. Top and side views of optimized structures of pure germanene with (a) $H_2S$, (b) $SO_2$, and (c) $CO_2$, monovacancy germanene with (d) $H_2S$, (e) $SO_2$, and (f) $CO_2$, divacancy germanene with (g) $H_2S$, (h) $SO_2$, and (i) $CO_2$, and trivacancy germanene with (d) $H_2S$, (e) $SO_2$, and (f) $CO_2$, respectively. Cyan, green, yellow, red and brown balls represent Ge, H, S, O, and H atoms, respectively.

A large adsorption energy of -2.58 eV is found in this configuration with a surface-gas distance of 2.04 Å. The adsorption energy $SO_2$ is almost the same as $H_2S$ and 1.7 times higher than M-germanene nanosheet. Though the bond length of S-O slightly elongated and O-S-O angle increased a little bit from 1.58 to 1.6 Å and 110.2 to 111.05°, respectively, compared to the pristine case. There is moderate improvement found for the case of $CO_2$ adsorbance on M-germanene as in Figure 2(f). It is reduced in this case which is from -0.42 to -0.49 eV and a binding distance of 3.61 Å is calculated in this case and the bond angle remains the same as pristine.

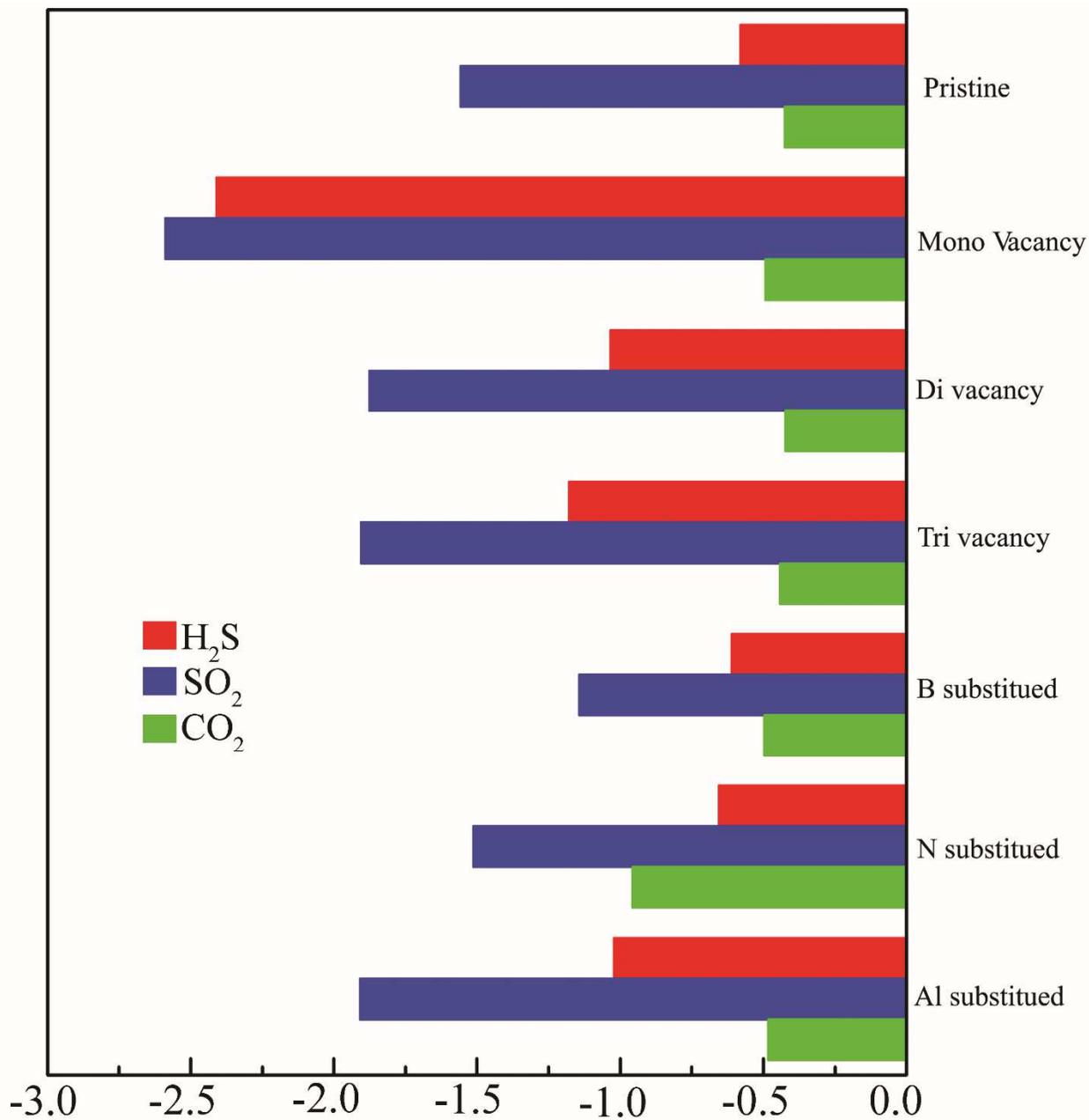

Figure 3. van der Waals induced adsorption energies of incident gases on pristine, defected, and substituted germanene.

While the addition of monovacancy in germanene nanosheet enhanced the adsorption sensitivity of the investigated gases, nonethelessthe divacancy in the system has proved to be less effective. Adsorption energies of -1.03, -1.87, and -0.42 eV have been calculated on D-germanene in the instance of $H_2S$, $SO_2$, and $CO_2$ as shown in Figure 2(g)-2(i), though they are still more attractive numbers compared to D-silicene [17] and GeH sensors [33]. It is observed that $H_2S$ adsorption increases by two times while $SO_2$ and $CO_2$ remain the same compared to their pristine cases. The relative surface-gas distances are found to be 2.56, 2.01 and 3.7 Å for $H_2S$, $SO_2$, and $CO_2$, respectively which are higher than those of monovacancy because of the adsorption energy difference. Finally, trivacancy defects have been considered for T-germanene and the most feasible structure are shown in Figure 2(j)-2(l). Those structures offer adsorption energies of -1.17, -1.9, and -0.44 eV for $H_2S$, $SO_2$, and $CO_2$, respectively. They also maintain a distance between gas and germanene atom which are of 2.58, 2, and 3.62 Å for $H_2S$, $SO_2$, and $CO_2$, respectively indicating similar adsorption behavior as D-germanene gas adsorption.

Additionally, impurity effects on the adsorption behavior of gas molecules have been investigated by replacing germanium atoms with foreign atoms in a germanene nanosheet. The dopants considered here include B, N, and Al atoms, which have already been replaced in the germanene sheet for diverse purposes such as tuning the electronic and magnetic properties [48, 49]. All of the stable structural configurations of the doped gas sensing system have been presented in supplementary information (Figure S1). In the case of $H_2S$ adsorption, it has been observed that B-doped and N-doped germanene do not cause significant changes in adsorption energies which are -0.61 and -0.65 eV, respectively while Al-doped germanene offers considerable improvement (-1.2 eV). The absolute values of adsorption energy decrease upon adsorption of $SO_2$ on B-doped (-1.14 eV) and N-doped germanene (-1.51 eV), in which the drop in value is more pronounced in the former case. However, Al dopant enhances the interaction of $SO_2$ with germanene. The adsorption energy of $SO_2$ on Al-doped germanene is -1.90 eV. Although, no significant changes have found upon adsorption of $CO_2$ on B-doped (-0.49 eV) and Al-doped (-0.48 eV) germanene, nonetheless, N-doped germanene offers a substantial improvement in adsorption of $CO_2$. The adsorption energy of $CO_2$ on N-doped germanene (-0.95 eV) is almost two times greater than that of pristine germanene. Adsorption energies calculated for defected and doped germanene-gas sensing system are presented in Figure 3 for comparison.

Table 1. Band gap and charge transfer of $H_2S$, SO2, and $CO_2$ gas molecules due to adsorption on germanene nanosheet.

| Gas | Structure | Adsorption Energy, eV | Bandgap, $E_g$ eV | Charge transfer, $e$ |
|---|---|---|---|---|
| $H_2S$ | P-Ge | -0.58 | 0.04 | -0.221 |
| | M-Ge | -2.41 | 0.17 | 0.036 |
| | D-Ge | -1.03 | 0.19 | -0.324 |
| | T-Ge | -1.18 | 0.11 | -0.329 |
| | B-Ge | -0.61 | 0.27 | -0.252 |
| | N-Ge | -0.66 | 0.18 | -0.134 |
| | Al-Ge | -1.02 | 0.16 | -0.373 |
| $SO_2$ | P-Ge | -1.56 | 0.14 | 0.107 |
| | M-Ge | -2.59 | 0.02 | 0.194 |
| | D-Ge | -1.88 | 0.18 | 0.185 |
| | T-Ge | -1.90 | 0.15 | 0.210 |
| | B-Ge | -1.14 | 0.33 | -0.046 |
| | N-Ge | -1.51 | 0.31 | 0.108 |
| | Al-Ge | -1.91 | 0.10 | 0.083 |
| $CO_2$ | P-Ge | -0.42 | 0.00 | -0.051 |
| | M-Ge | -0.49 | 0.11 | -0.054 |
| | D-Ge | -0.42 | 0.25 | -0.044 |
| | T-Ge | -0.44 | 0.13 | -0.042 |
| | B-Ge | -0.50 | 0.04 | -0.038 |
| | N-Ge | -0.96 | 0.22 | 0.279 |
| | Al-Ge | -0.48 | 0.06 | -0.090 |

An electronic charge transfer occurs when gas molecules adsorb on germanene nanosheet. The amount of charge transfer between surface and gas molecules have been calculated using Mulliken population analysis [50] and is summarized in Table 1 for the most stable configurations considered here. The positive value of charge transfer indicates gas molecules gain charges from the surface while negative values means the surface receives charges from gases. $H_2S$ and $CO_2$ gas molecules are electron-donating gas molecules whereas $SO_2$ has the electron extracting capability. From the Table 1, it can be seen that $H_2S$ donates a significant amount of charge to the D-Ge (-0.324$e$), T-Ge (-0.329$e$), B-Ge (-0.252$e$) and Al-Ge (-0.373$e$) system. $H_2S$ molecules show strong adsorption in monovacancy defected germanene where a tiny amount of charge transfer occurs due to one of the H atoms dislodging itself from H-S-H bond and adsorbed strongly to make a bond with the germanium atom on the nanosheet, as shown in Figure 2(d). Hence, dissociated H atom gains charges of 0.099$e$ due to the higher electronegativity of hydrogen atom than germanium and deserted H-S bond loses charge in this process.

Being an electron donating character, $CO_2$ adsorption on germanene nanosheet shows similar charge transfer characteristics as $H_2S$. Compared to pristine germanene, it has been seen that vacancy defects and doping do not change much in the adsorption energy or the charge transfer of $CO_2$ except in the case of N-doped germanene, where a strong adsorption energy of -0.96 eV achieved with a large charge transfer of 0.279$e$. The most stable configurations for each case are shown in Figure S1. Though the charge transfer is expected to be negative, the gas molecule being adsorbed on nitrogen gains charges because of higher electronegativity of nitrogen compared to germanium. Also, one of the oxygen atoms of $CO_2$ makes a bond with a germanium atom on the surface to support the idea of charge gain on it. The same occurrence happens again because of C-Ge and O-Ge bond creation on the M-Ge nanosheet upon $CO_2$ adsorption, as shown in Figure 2(f). This process transfers 0.191$e$ charge to the gas molecule as carbon and oxygen are more electronegative compared to germanium atoms on the surface. Finally, $SO_2$ molecules transfer charges to the germanene surface upon adsorption which are 0.107$e$, 0.194$e$, 0.185$e$, 0.210$e$, and 0.108$e$ for P-Ge, M-Ge, D-Ge, T-Ge and N-Ge systems respectively.

The origin of physisorption and chemisorption can be understood further using the density of states (DOS) analysis of surface-adsorbate systems. Figure 4 presents the DOS of the pristine germanene-gas systems. It can be seen that the highest occupied molecular orbital (HOMO) and the lowest unoccupied molecular orbital (LUMO) of $H_2S$ and $CO_2$ are located away from the Dirac point (Fermi level). Therefore, the Dirac cone properties of germanene remain almost unaffected when $CO_2$ and $H_2S$ molecules are adsorbed. Therefore, they are weakly physisorbed on germanene, approaching the case of $H_2S$ and $CO_2$ adsorbed on silicene [17]. However, the changes are more pronounced for $H_2S$ adsorption. Quite the reverse, $SO_2$ gas molecule adsorption on pristine germanene nanosheet greatly influence the electronic properties. The resulting structural deformation induces symmetry breaking vanishing the Dirac cone completely through strong Ge-O bond illuminating that the $SO_2$ is strongly chemisorbed. As previously studied that the atoms in the germanene layer are offset from the atomic plane, the typically perpendicular $p_z$ orbital is gradually interacting with the $sp^2$ orbitals due to the reduced angle between them resulting in a mixed-phase material that is predominantly $sp^2$ with a small amount of $sp^3$ character present [10]. This phenomenon makes germanene more reactive than graphene to the gas molecules.

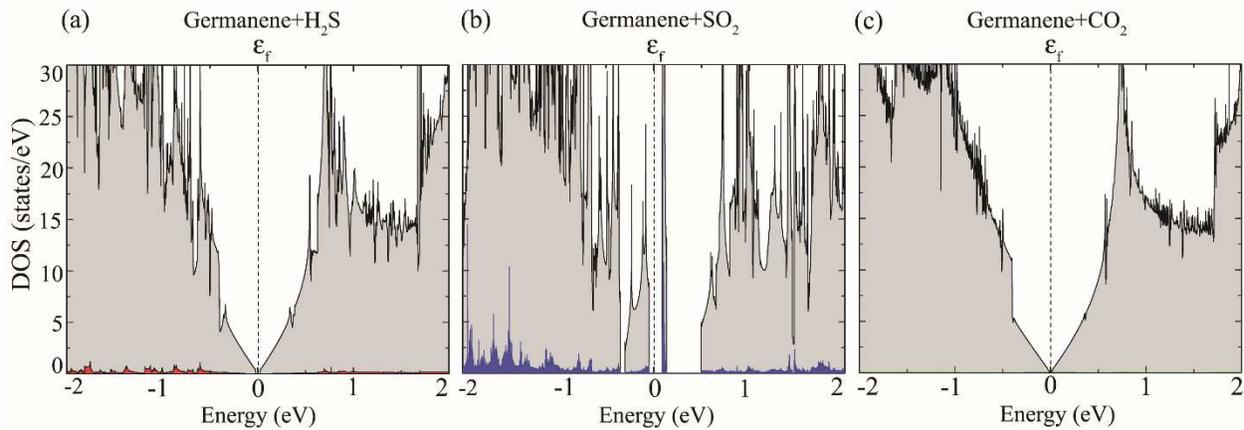

Figure 4. The density of states (DOS) of the pristine germanene nanosheet adsorbed by (a) $H_2S$ and (b) $SO_2$ and (c) $CO_2$ molecule. The grey shaded area represents the total DOS, while the red, blue, and green colors account for the projection of DOS onto $H_2S$, $SO_2$, and $CO_2$ molecule, respectively. Here the Fermi level is shifted to zero.

Additionally, it can be seen that some states are near the Fermi level indicating p-type doping of germanene by $SO_2$ adsorption on pristine germanene sheet, which agrees well with the positive charge transfer on $SO_2$ upon interaction with pristine germanene. Due to strong chemisorption of $SO_2$ on germanene sheets make them a promising candidate for future $SO_2$ gas sensing device.

DOS can provide detail insight in response to higher adsorption energies enhanced by vacancy defects in germanene as shown in Figure 5. Dissimilar to P-germanene, adsorption of gas molecules on M-germanene leads to a significant influence in the DOS because of strong S-Ge, C-Ge, and Ge-H bonds in $SO_2$, $CO_2$, and $H_2S$, respectively. The bond strength is believed to be enhanced due to the intensive hybridization of C-Ge and S-Ge bonds. Moreover, M-germanene turns to semiconducting upon gas molecules adsorption and single point vacancy defect results in localized states to cross bands linearly at *K*-point to open the band gap due to symmetry breaking [51].

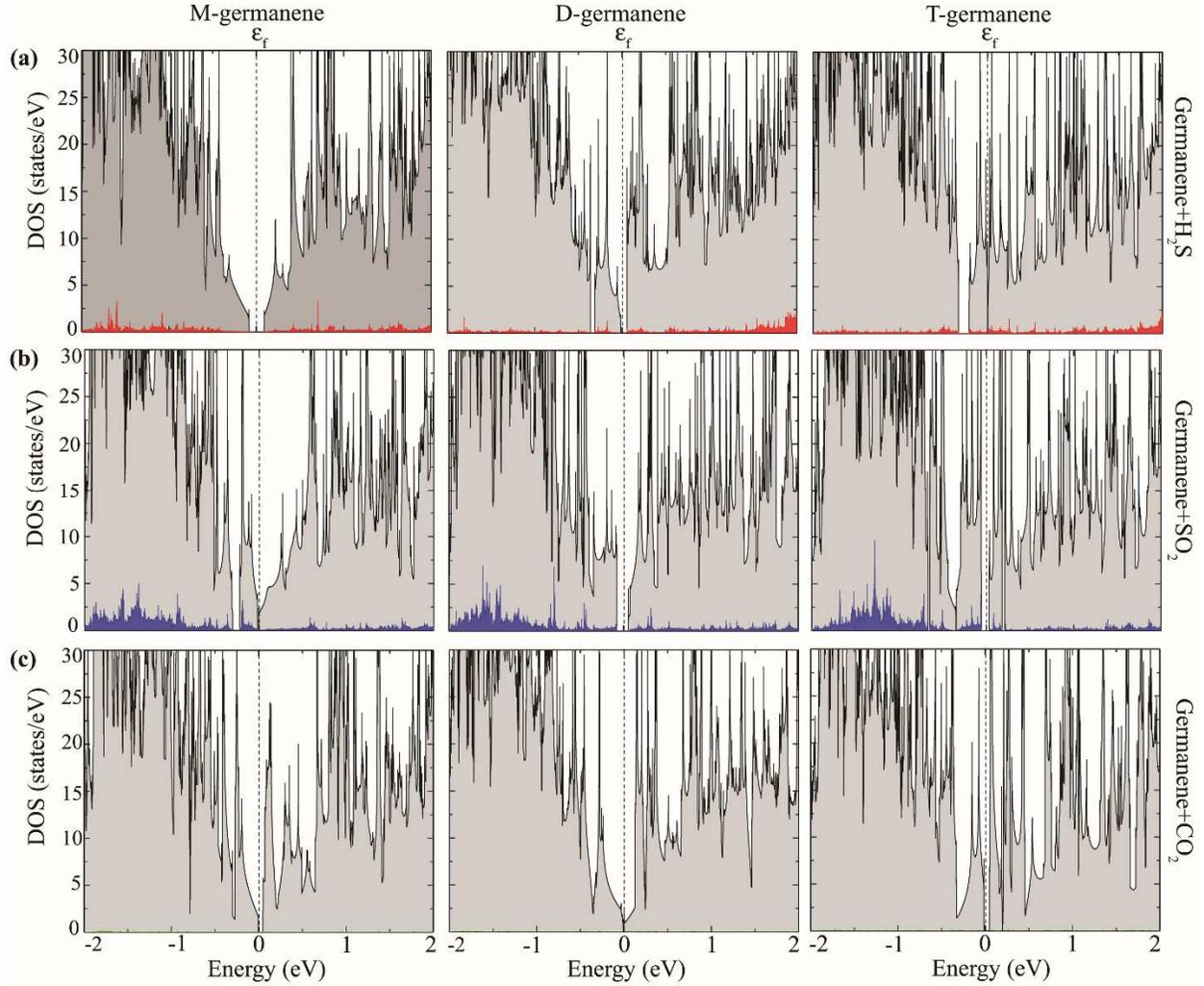

Figure 5. The density of states (DOS) of the vacancy defected germanene nanosheet adsorbed by (a) $H_2S$ and (b) $SO_2$ and (c) $CO_2$ molecule. The grey shaded area represents the total DOS, while the red, blue, and green colors account for the projection of DOS onto $H_2S$, $SO_2$, and $CO_2$ molecule, respectively. Here the Fermi level is shifted to zero.

In the meantime, D-germanene exhibits reduced adsorption energies upon gas molecules adsorption which could be due to less reactivity of germanene atoms in this structure compared to M-germanene. Still, there is no formation of the ring-like structure as in D-silicene in the Ref. [17], and this is why D-germanene exhibits strong adsorption for $H_2S$ and $SO_2$ as compared to P-germanene. From the DOS analysis in Figure 5(c), it is easily discernible that $CO_2$ shows a smaller impact on the total DOS of germanene-$CO_2$ system. On the contrary, $H_2S$ located at the position of the removed Ge atoms provide significant hybridization of S-Ge and H-Ge near the Fermi level. Interestingly, $SO_2$ forms a strong bond through Ge-O to provide strong adsorption similar to P-germanene. Finally, trivacancy does not show any significant improvement in adsorption energies over D-germanene for $H_2S$, $SO_2$, and $CO_2$.

In summary, it is worth mentioning that P-germanene is reasonably sensitive to $H_2S$ while $SO_2$ shows strong adsorption, and $CO_2$ has comparatively weak adsorption. The recovery time of the gas sensing material is the parameter to evaluate the performance of a gas sensor [29]. The recovery time $\tau$ can be calculated using the conventional transition state theory describe by the following formula,

$$\tau \propto v_0^{-1} \exp(-E_{ad} / k_B T) \tag{5}$$

Here, $v_0$ is the attempt frequency. The relation established by the equation shows that the larger the adsorption energy (more negative) the longer the recovery time required for the particular sensor. Adsorption energies for recoverable gas sensing device are moderate because of reasonable recovery time in the process of desorption. Meanwhile, the defected germanene make it useful for disposable gas sensing devices due to strong adsorption energies. Although $CO_2$ is weakly adsorbed on defected germanene with almost same adsorption energy as P-germanene, it shows moderate adsorption on N-doped germanene, making it suitable for reusable sensing material.

## 4. Conclusion

It was found that $CO_2$ gas molecules shows weak adsorption while $H_2S$ is moderate and $SO_2$ shows strong adsorption on pristine germanene nanosheet. It has also been investigated that defects and doping in germanene change the respective electronic properties and the adsorption characteristics significantly. M-silicene shows a dramatic increase in adsorption energies of $H_2S$ and $SO_2$ which are 4 times and 1.66 times higher compared to pristine germanene nanosheet. Increasing defects does not improve the adsorption over M-germanene while T-germanene is the least effective case. Strong chemisorption is observed in the case of $SO_2$ with vacancy defect or doping in a germanene nanosheet. The electronic properties portrayed by the density of states of the germanene nanosheet system with different defects and substitutions are changed as a consequence of the charge-transfer mechanism between germanene-gas molecules. The doping of germanene with B, N, and Al dopants does not seem to alter the adsorption energy for $H_2S$ and $SO_2$, but for N-doped germanene, the adsorption of $CO_2$ increases a lot. Finally, it could be concluded that defects and doping in germanene increase the adsorption with a cost of sensing devices to be irreversible.